\documentclass[preprintnumbers,nofootinbib,a4paper,superscriptaddress]{revtex4-1}

\usepackage{amsmath,latexsym,amssymb,amsfonts}
\usepackage{float, graphicx}
\usepackage{epstopdf}
\usepackage{subfigure}
\usepackage{bm}
\usepackage{color}
\usepackage{hyperref}
\usepackage[dvips]{psfrag}

\usepackage{soul}  
\usepackage{xcolor}
\setstcolor{red}

\begin{document}

\title{Wheeler-DeWitt equation beyond the cosmological horizon: Annihilation to nothing, infinity avoidance, and loss of quantum coherence}

\author{Chen-Hsu Chien} \email[]{chenhsu0223@gmail.com}
\affiliation{Institute of Physics, Academia Sinica, Taipei 11529, Taiwan}
\author{Gansukh Tumurtushaa} \email[]{gansukh@jejunu.ac.kr}
\affiliation{Department of Science Education, Jeju National University, Jeju, 63243, Korea}
\author{Dong-han Yeom} \email[]{innocent.yeom@gmail.com}
\affiliation{Department of Physics Education, Pusan National University, Busan 46241, Republic of Korea}
\affiliation{Research Center for Dielectric and Advanced Matter Physics, Pusan National University, Busan 46241, Republic of Korea}
\affiliation{Leung Center for Cosmology and Particle Astrophysics, National Taiwan University, Taipei 10617, Taiwan}
\date{\today}

\begin{abstract}
We investigate the Schwarzschild-(anti) de Sitter spacetime with the anisotropic metric ansatz. The Wheeler-DeWitt equation for such a metric is solved numerically. In the presence of the cosmological constant $\Lambda$, we show that two classical wave packets can be annihilated inside the black hole horizon, i.e., the \textit{annihilation-to-nothing} scenario. It is interesting that the Wheeler-DeWitt equation can be extended to the asymptotic de Sitter spacetime outside the cosmological horizon. Surprisingly, the only bounded nontrivial wave function beyond the cosmological horizon satisfies the DeWitt boundary condition, i.e., the wave function must vanish at a certain finite radius. This might be an alternative explanation to the classicalization of quantum fluctuations in the de Sitter space, where this topic is also related to decoherence.

\end{abstract}

\maketitle

\section{Introduction}\label{sec:intro}
Black hole physics has been an active topic in modern theoretical physics for the past few decades. {\color{black} However,} it suffers from the spacetime defect, the singularity at its core~\cite{Hawking:1970zqf}, and the information loss paradox because of the lack of fundamental quantum gravity theory~\cite{Hawking:1976ra}. One of the most conservative approaches is to solve the quantum Hamiltonian constraint equation, the Wheeler-DeWitt equation, with a specific metric ansatz~\cite{Cavaglia:1994yc}. The wave function solution could be essential for understanding quantum gravity and potentially address these problems.

In previous works, the Wheeler-DeWitt equation for the interior Schwarzschild black hole can be simplified by the anisotropic model (the Kantowski-Sachs metric)~\cite{Kantowski:1966te}. The partial differential equation with two canonical variables can be solved by separating variables. The bounded wave function with suitable assumptions presents a new interpretation, \textit{annihilation-to-nothing}~\cite{Bouhmadi-Lopez:2019kkt}. The wave function shows a quantum bounce at half the black hole size. The DeWitt boundary condition, vanishing wave function, within \textit{annihilation-to-nothing} has been discussed in Ref.~\cite{Brahma:2021xjy}. By introducing the spinorial Wheeler-DeWitt equation for the interior of a higher-dimensional planar topological black hole metric, the alternative interpretation could be “no annihilation” or “annihilation-to-something”~\cite{Kan:2022ism}. Therefore, the quantum behavior around half of the black hole size is worth additional attention due to these wave function solutions.

From the observational point of view, the standard $\Lambda$CDM model of cosmology implies that our universe has a nonzero positive cosmological constant $\Lambda$~\cite{Planck:2018vyg}. On the theoretical physics side, anti-de Sitter spacetime and conformal field theory (AdS/CFT) are crucial for quantum gravity theory~\cite{Aharony:1999ti}. For more generic cases, we consider the Schwarzschild-(anti) de Sitter metric, the static black hole solutions in the (anti) de Sitter spacetime, throughout this work. The mass and entropy for the Schwarzschild-(anti) de Sitter spacetime in the Wheeler-DeWitt equation has been discussed in Ref.~\cite{Lopez-Dominguez:2011tcj}. The Wheeler-DeWitt equation for the Schwarzschild-(anti) de Sitter metric is sophisticated; therefore, we solve it numerically. For sufficiently small $\Lambda$, i.e. $\Lambda\sim\pm\mathcal{O}(10^{-9})$, we show that $\mathit{annihilation}$-$\mathit{to}$-$\mathit{nothing}$ is a generic feature. Moreover, in de Sitter spacetime, the wave function outside the cosmological horizon also vanishes at the effective potential boundary where the geometry should be classical.

This paper is organized as follows. In Sec.~\ref{sec:WF}, we derive the Wheeler-DeWitt equation for the Schwarzschild-(anti) de Sitter metric. In Sec.~\ref{sec:NA}, we numerically solve the equation with Gaussian wave packets and discuss the valid regions separately. Finally, in Sec.~\ref{sec:con}, we summarize and point out possible future applications of the framework.

\section{The Wheeler-DeWitt equation of Schwarzschild-(anti) de Sitter metric}\label{sec:WF}

The interior Schwarzschild-(anti) de Sitter metric takes the form of
\begin{eqnarray}\label{eq:sbh}
ds^2= -\left(\frac{\Lambda}{3}t^2+\frac{r_s}{t}-1\right)^{-1} dt^2+\left(\frac{\Lambda}{3}t^2+\frac{r_s}{t}-1\right) dR^2+t^2d\Omega^2,
\end{eqnarray}
where $r_s=2M$ is the Schwarzschild radius and $\Lambda$ is the cosmological constant.
When $\Lambda<0$, the Schwarzschild-anti-de Sitter metric has a black hole horizon in which Eq.~(\ref{eq:sbh}) is valid inside this horizon. When $\Lambda>0$, the Schwarzschild-de Sitter metric has two horizons in which it is valid both inside the black hole horizon and outside the cosmological horizon~\cite{Stuchlik:1999qk}.\footnote{In de Sitter spacetime, the $\Lambda$ has an upper bound, the Nariai limit $\Lambda\le4/9$ for $r_s=1$, to preserve the horizons~\cite{Nariai}.}

Under particular diffeomorphism, this metric can transform into the anisotropic model~\cite{Kantowski:1966te}
\begin{eqnarray}\label{eq:ani}
ds^2= -{N^2}(t)dt^2+{a^2}(t)dR^2+r_s^2 \frac{{b^2}(t)}{{a^2}(t)}d\Omega_2^2,
\end{eqnarray}
where $N(t)$ is the lapse function, $a(t)$ and $b(t)$ denote two canonical dimensionless variables. For simplicity, we define $X\equiv\ln{a}$ and $Y\equiv\ln{b}$. 
The classical on-shell trajectory in the $(X,Y)$ space is
\begin{eqnarray}\label{eq:XY}
e^X+e^{-X}=e^{-Y}+\frac{\Lambda}{3}r_s^2 e^{2Y-3X}.
\end{eqnarray}
The Wheeler-DeWitt equation for Schwarzschild-(anti) de Sitter black hole can be further derived following Ref.~\cite{Bouhmadi-Lopez:2019kkt} as
\begin{eqnarray} \label{eq:wdw}
\left[\frac{\partial^2}{\partial X^2}-\frac{\partial^2}{\partial Y^2}-V(X,Y)\right] \Psi(X,Y)=0,
\end{eqnarray}
where the potential is
\begin{eqnarray} \label{eq:V}
V(X,Y)=-4r_s^2e^{2Y}+4 r_s^4\Lambda e^{4Y-2X}.
\end{eqnarray}
In the limit, $\Lambda\rightarrow 0$, the Wheeler-DeWitt equation in Ref.~\cite{Bouhmadi-Lopez:2019kkt} is recovered. The sign in front of the potential is defined in Appendix~\ref{appendix:unb}.

\begin{figure}[h]
\begin{center}
    \includegraphics[width=0.45\textwidth]{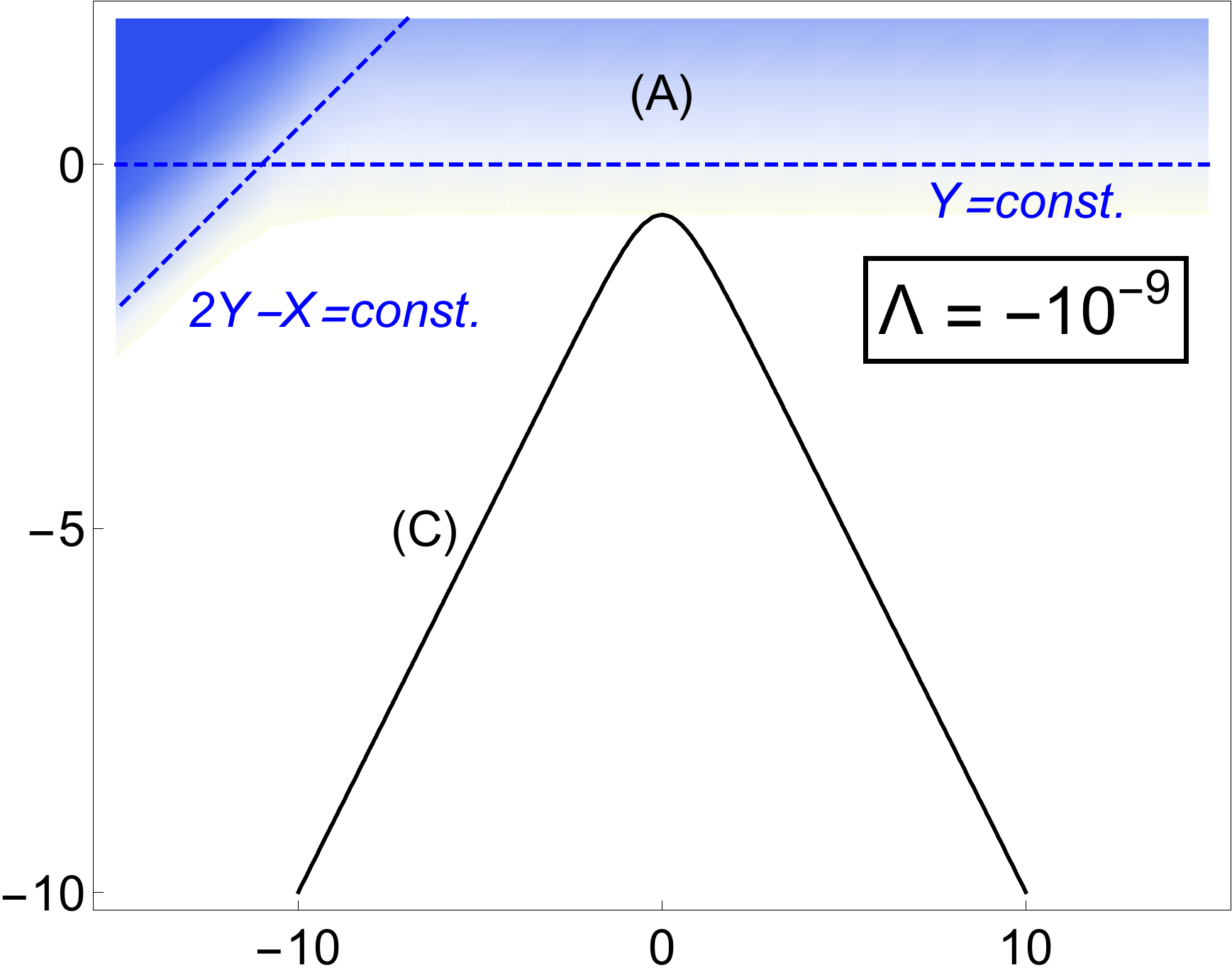}\hspace{0.05 cm}
    \includegraphics[width=0.45\textwidth]{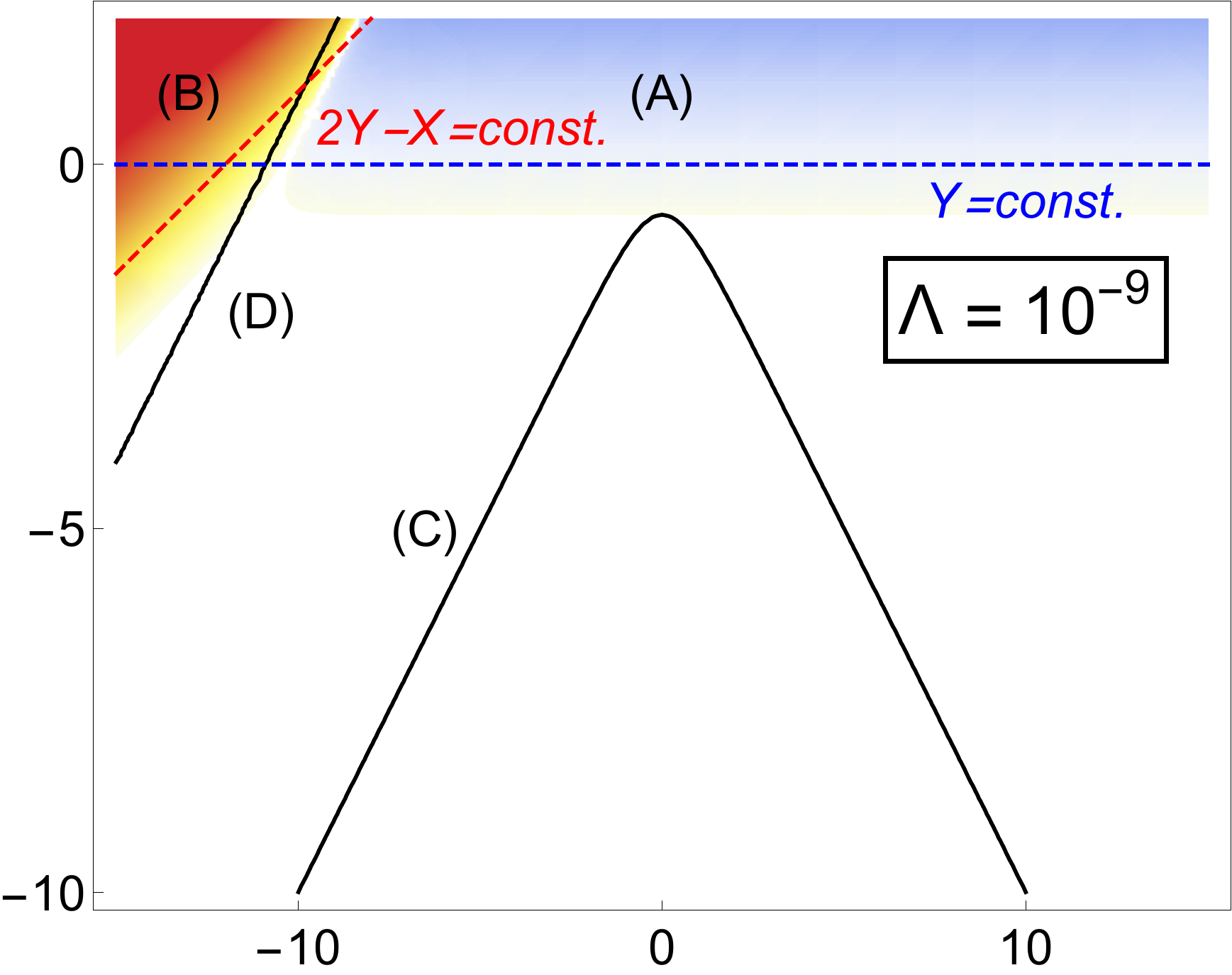}
    \caption{The anti-de Sitter spacetime with $\Lambda=-10^{-9}$ and the de Sitter spacetime with $\Lambda=10^{-9}$, left and right, respectively. The region (A) is the potential hollow and (B) is the potential barrier, and the curve (C) is the classical on-shell trajectory Eq.~(\ref{eq:XY}) inside the black hole horizon and (D) is the classical on-shell trajectory outside the cosmological horizon. The warm (cold) color region denotes the positive (negative) potential region. The potential value $\vert V\vert$ increases from lighter to darker.} 
    \label{fig:00}
\end{center}
\end{figure}

The potential is highly correlated with the sign of $\Lambda$; see Fig.~\ref{fig:00}. In $\Lambda=0$ spacetime, the potential decrease exponentially in $+Y$ direction and behaves like a potential hollow $V\rightarrow -\infty$ beyond the effective potential boundary $Y=constant$. In anti-de Sitter spacetime $\Lambda<0$, since the identical sign within the potential, the potential hollow has enlarged proportionally to $2Y-X=constant$. In de Sitter spacetime $\Lambda>0$, because of the different signs, the potential hollow has dwindled, and a potential barrier $V\rightarrow \infty$ appears proportionally to $2Y-X=constant$. The potential hollow and potential barrier are shown as (A) and (B) in Fig.~\ref{fig:00} and its corresponding effective potential boundaries are shown on the left of Fig.~\ref{fig:01}.~\footnote{The boundary value is chosen artificially where the wave function starts to behave differently. \color{black}{The effective potential boundary shifts corresponding to different values of $\sigma$.}}

\begin{figure}[h]
    \centering
    \includegraphics[width=0.45\textwidth]{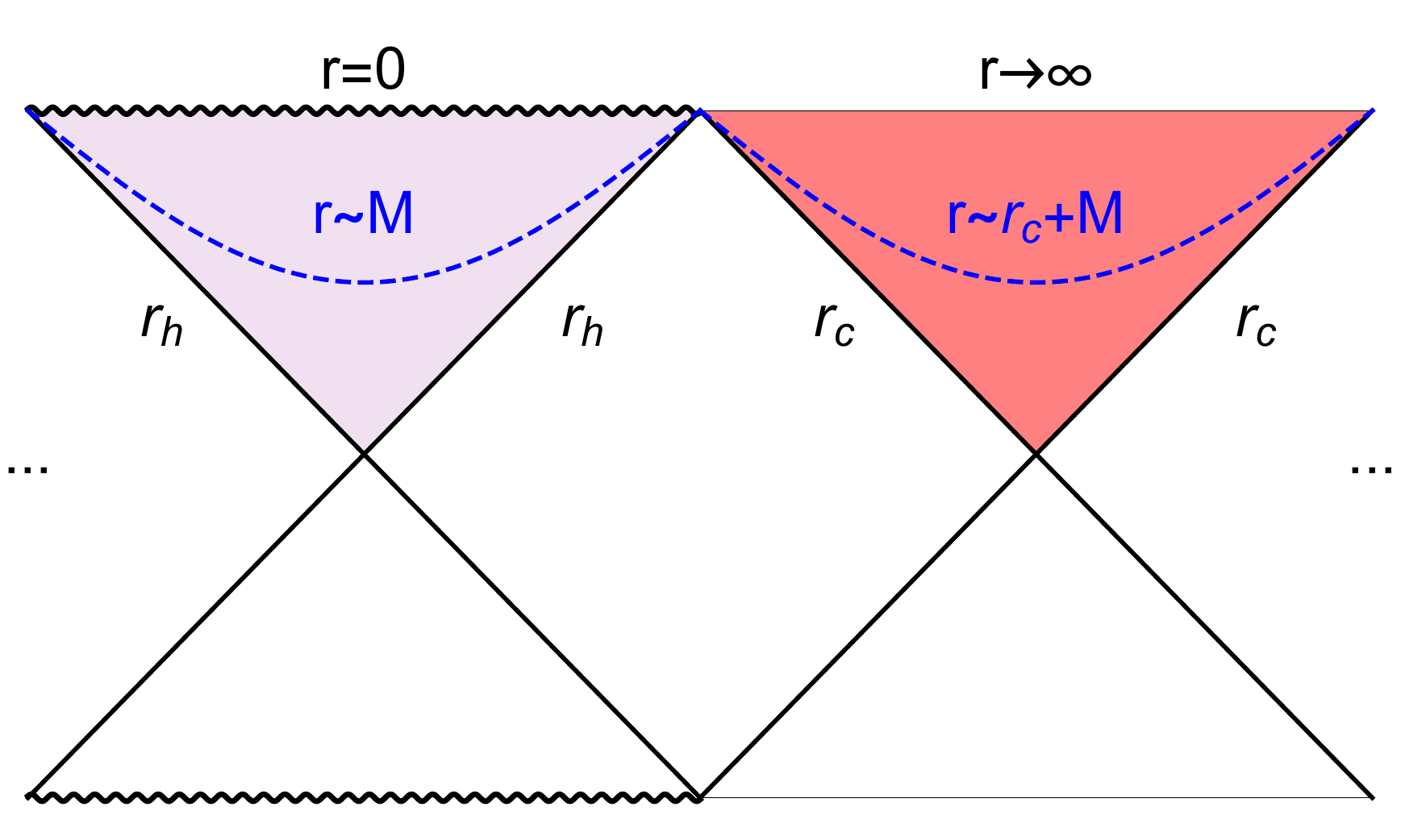}
    \caption{The Penrose diagram of the Schwarzchild-de Sitter black hole. The $r_h$ denotes the black hole horizon, while the $r_c$ is the cosmological horizon. The singularity is at $r=0$ whereas $r\rightarrow \infty$ is the asymptotic infinity. The shaded triangles are the regions simulated by the anisotropic model in Eq.~(\ref{eq:ani}), where the purple triangle represents the region inside $r_h$ and the pink one denotes the region outside $r_c$.}
    \label{fig:00PD}
\end{figure}

The Penrose diagram of the Schwarzchild-de Sitter black hole is shown in Fig.~\ref{fig:00PD}. Inside the black hole horizon for $\Lambda<0$ and $\Lambda=0$, it is also similar to the purple triangle~\footnote{The Penrose diagram inside the black hole horizon is also similar to Fig. 1 and 2 in Ref~\cite{Bouhmadi-Lopez:2019kkt}.\\The Penrose diagram of the Schwarzchild-de Sitter black hole at the Nariai limit is not considered here. Such diagrams can be found in Refs.~\cite{Hartong:2004rra}.} and that for $\Lambda>0$ outside the cosmological horizon is the pink triangle in Fig.~\ref{fig:00PD}. The Penrose diagram of Schwarzchild (anti)-de Sitter black hole is fully analyzed in Refs.~\cite{Hartong:2004rra}.

\section{Numerical Analysis}\label{sec:NA}
Since the potential depends on both $X$ and $Y$, the Wheeler-DeWitt equation Eq.~(\ref{eq:wdw}) cannot be solved simply by separating variables; therefore, we use the numerical approach. In the following subsections, we investigate the wave function inside the black hole horizon (the purple triangle) for $\Lambda<0$, $\Lambda=0$, and $\Lambda>0$, and outside the cosmological horizon (the pink triangle) for $\Lambda>0$, the special case, separately.
\subsection{Inside the black hole horizon}\label{sec:IN}
To solve the Wheeler-DeWitt equation, one must impose boundary conditions. As suggested in Ref.~\cite{Bouhmadi-Lopez:2019kkt}, we first impose two wave pulses at the horizon and the singularity.

The boundary condition of the integration domain $\left(X_L, X_R\right)$ × $\left(-Y_m, Y_M\right)$ is given as follows:

\begin{itemize}
    \item[-] For $\Psi (X, -Y_m)$, a localized (e.g., Gaussian) wave packet must be located at the on-shell, i.e., $(X_L,-Y_m)$ and $(X_R,-Y_m)$ are points on Eq.~(\ref{eq:XY}). The wave packet at $Y=-Y_m$ can be chosen to be
    \begin{eqnarray} \label{eq:wfbc}
    \Psi(X,-Y_m)&=&\frac{A}{(2\pi\sigma^2)^{1/4}}\left(e^{-\frac{(X-X_L)^2}{4\sigma^2}}-e^{-\frac{(X-X_R)^2}{4\sigma^2}}\right),
    \end{eqnarray}
    where $A$ and $\sigma$ are constants. Here, to annihilate each other, two Gaussian pulses have opposite amplitudes.
    \item[-] For $\partial_{Y} \Psi (X, -Y_m)$, we consider that the left pulse at $(X_L,-Y_m)$ goes $X$-increasing direction, while the right pulse $(X_R,-Y_m)$ goes $X$-decreasing direction. The corresponding condition is
    \begin{eqnarray} \label{eq:pwfbc}
    \partial_{Y}\Psi(X,-Y_m)&=&\frac{A}{2\sigma^2(2\pi\sigma^2)^{1/4}}\left((X-X_{L})e^{-\frac{(X-X_L)^2}{4\sigma^2}}-(X_R-X)e^{-\frac{(X-X_R)^2}{4\sigma^2}}\right).
    \end{eqnarray}
    \item[-] For $\Psi (X_R/X_L, Y)$ and $\partial_{X}\Psi (X_R/X_L, Y)$, as long as the standard deviation of each Gaussian wave packet is sufficiently small, one can choose small value close to 0.
\end{itemize}

\begin{figure}[!ht]
    \centering
    \includegraphics[width=0.45\textwidth]{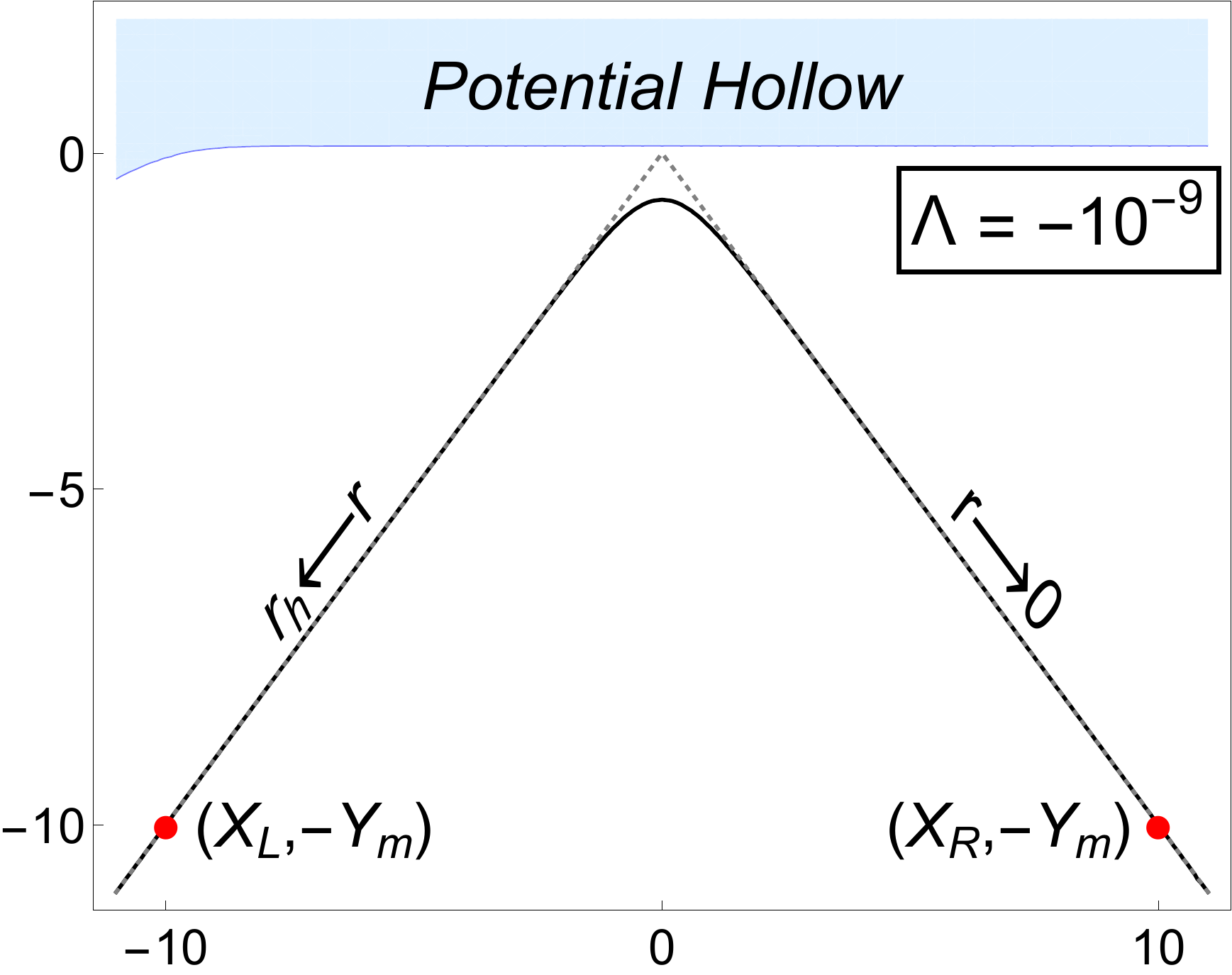}\hspace{0.05 cm}
    \includegraphics[width=0.45\textwidth]{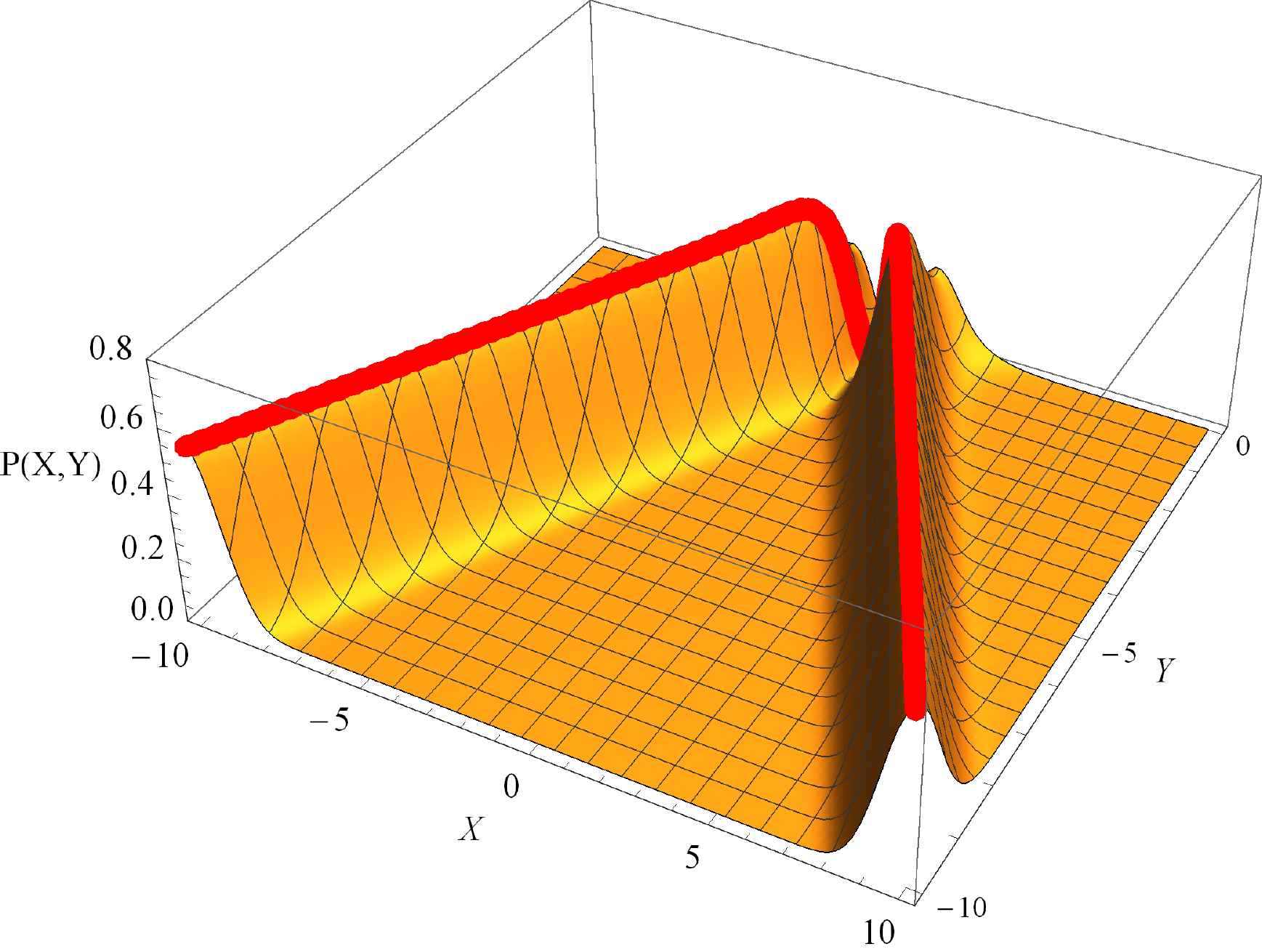}
    \includegraphics[width=0.45\textwidth]{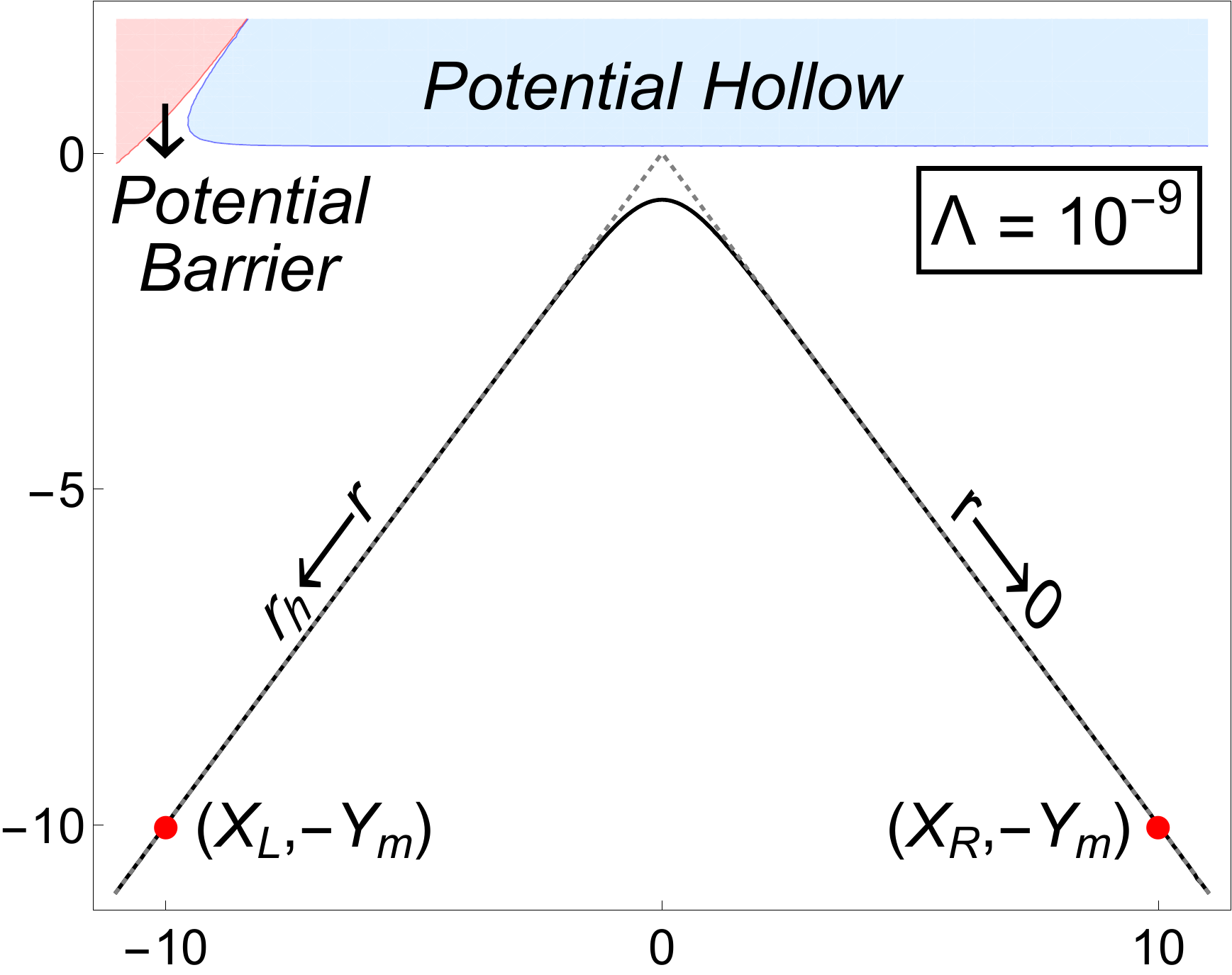}\hspace{0.05 cm}
    \includegraphics[width=0.45\textwidth]{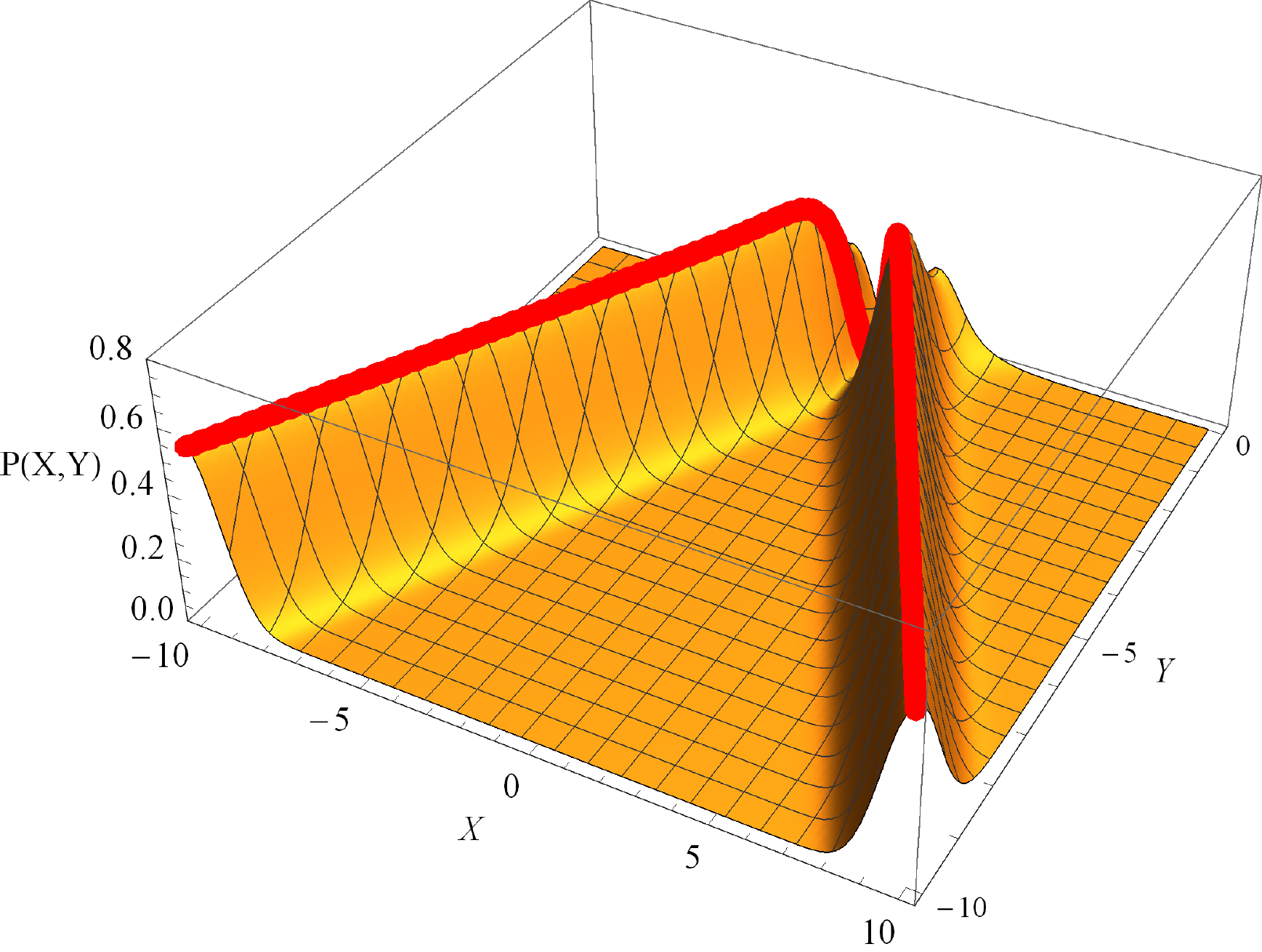}
    \caption{Left: The black line is the classical on-shell trajectory Eq.~(\ref{eq:XY}) and curve (C) in Fig.~(\ref{fig:00}). The blue and red regions are the effective potential hollow and the effective potential barrier with their boundary values $V\leq-5$ and $V\geq 5$. The effective potential boundary value $\vert V\vert=5$ is chosen artificially where the wave function starts evolving. Right: The modulus squared of the wave function with certain values of the cosmological constant are shown numerically with $r_s=A=1$ and $\sigma=0.75$. Here, the domain is set $\left(X_L, X_R\right)$ × $\left(-Y_m, Y_M\right)=\left(-10, 10\right)$ × $\left(-10, 0\right)$. The incoming pulses from $X,$ $Y\rightarrow -\infty$ (the black hole horizon $r_h$) and from $X\rightarrow \infty,$ $Y\rightarrow -\infty$ (the singularity) can be annihilated around $X \sim 0$ ($r \sim M$).
    } 
    \label{fig:01}
\end{figure} 

In Fig.~\ref{fig:01}, the steepest-descent~\footnote{The concept of the steepest-descent is explained in Sec.~III in Ref.~\cite{Bouhmadi-Lopez:2019kkt}.} located on the classical on-shell trajectory Eq.~(\ref{eq:XY}). For sufficiently small $\Lambda$, since the classical on-shell trajectory does not cross the effective potential boundaries, two classical wave packets can be annihilated, i.e., the \textit{annihilation-to-nothing} scenario; also, the DeWitt boundary condition is satisfied at $X\sim0$~\cite{Bouhmadi-Lopez:2019kkt,Brahma:2021xjy}~\footnote{Although the classical on-shell trajectory still does not cross the effective potential boundary for $\Lambda\sim\pm\mathcal{O}(10^{-6})$, the numerical error in the modulus squared of the wave function would not be neglected. To avoid the numerical error, we consider $\Lambda\sim\pm\mathcal{O}(10^{-9})$ in this work.}. The wave function apparently blows up for $Y_M>0$, but this is due to numerical errors. 
{\color{black} One can control the divergent behavior by carefully tuning the boundary condition, as seen in Appendix~\ref{appendix:unb}. Thus, we believe that a bounded wave function is allowed.}

To summarize, the wave packets are completely annihilated at $X \sim 0$ ($r \sim M$), and the \textit{annihilation-to-nothing} scenario can be reproduced. {\color{black}This is not surprising because the analytic solution exists for $\Lambda = 0$~\cite{Bouhmadi-Lopez:2019kkt} and, as long as $|\Lambda|$ is sufficiently small, there must exist a corresponding bounded solution.} 
The contribution of $\Lambda$ is that the trajectory is shifted in Eq.~(\ref{eq:XY}).

\subsection{Outside the cosmological horizon}\label{sec:OT}
The Schwarzschild-de Sitter black hole has two horizons, the black hole horizon $r_h$ and the cosmological horizon $r_c$. The classical on-shell trajectory Eq.~(\ref{eq:XY}) outside the cosmological horizon, (D) in Fig.~\ref{fig:00}, can be analyzed similarly. 

The boundary conditions differ from those in Sec.~\ref{sec:IN}; therefore, we use a tilde to distinguish them. The boundary condition of the integration domain $\left(\widetilde{X}_L, \widetilde{X}_R\right)$ × $\left(-\widetilde{Y}_m, \widetilde{Y}_M\right)$ is given as follows:
\begin{itemize}
    \item[-] For $\Psi (X, -\widetilde{Y}_{m})$, a localized (e.g., Gaussian) wave packet must be located at the on-shell, i.e., $(\widetilde{X}_{L},-\widetilde{Y}_{m})$ is a point on Eq.~(\ref{eq:XY}). The wave packet at this boundary can be chosen to be
    \begin{eqnarray} \label{eq:wfbco}
    \Psi(X,-\widetilde{Y}_{m})&=&\frac{A}{(2\pi\sigma^2)^{1/4}}e^{-\frac{(X-\widetilde{X}_{L})^2}{4\sigma^2}}
    \end{eqnarray}
    where $A$ and $\sigma$ are constants. 
    \item[-] For $\partial_{Y} \Psi (X, -\widetilde{Y}_{m})$, we consider the pulse going $X$-increasing direction,
    \begin{eqnarray} \label{eq:pwfbco}
    \partial_{Y}\Psi(X,-\widetilde{Y}_{m})&=&\frac{A}{2\sigma^2(2\pi\sigma^2)^{1/4}}(X-\widetilde{X}_{L})e^{-\frac{(X-\widetilde{X}_{L})^2}{4\sigma^2}}
    \end{eqnarray}
    \item[-] One may find an endpoint on Eq.~(\ref{eq:XY}), say $(\widetilde{X}_{R},\widetilde{Y}_{m})$. With the foresight that it should vanish at the effective potential barrier boundary, $\Psi (\widetilde{X}_{R}, Y)$ and $\partial_{X}\Psi (\widetilde{X}_{R}, Y)$ are chosen to be zero there. For $\Psi (\widetilde{X}_{L}, Y)$ and $\partial_{X}\Psi (\widetilde{X}_{L}, Y)$, as long as the standard deviation of each Gaussian wave packet is sufficiently small, one can choose zero.
\end{itemize}

\begin{figure}[!ht]
    \centering
    \includegraphics[width=0.45\textwidth]{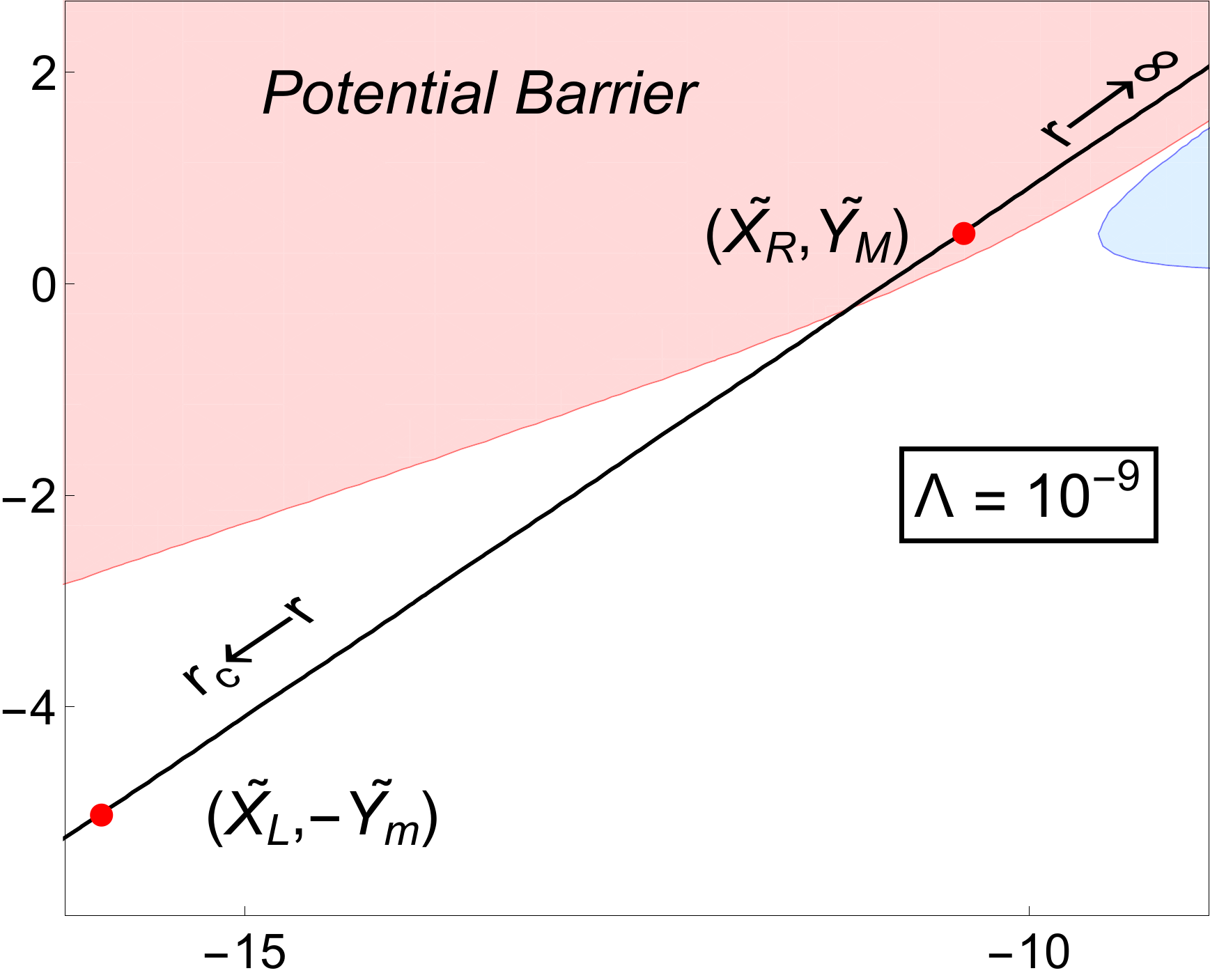}\hspace{0.05 cm}
    \includegraphics[width=0.45\textwidth]{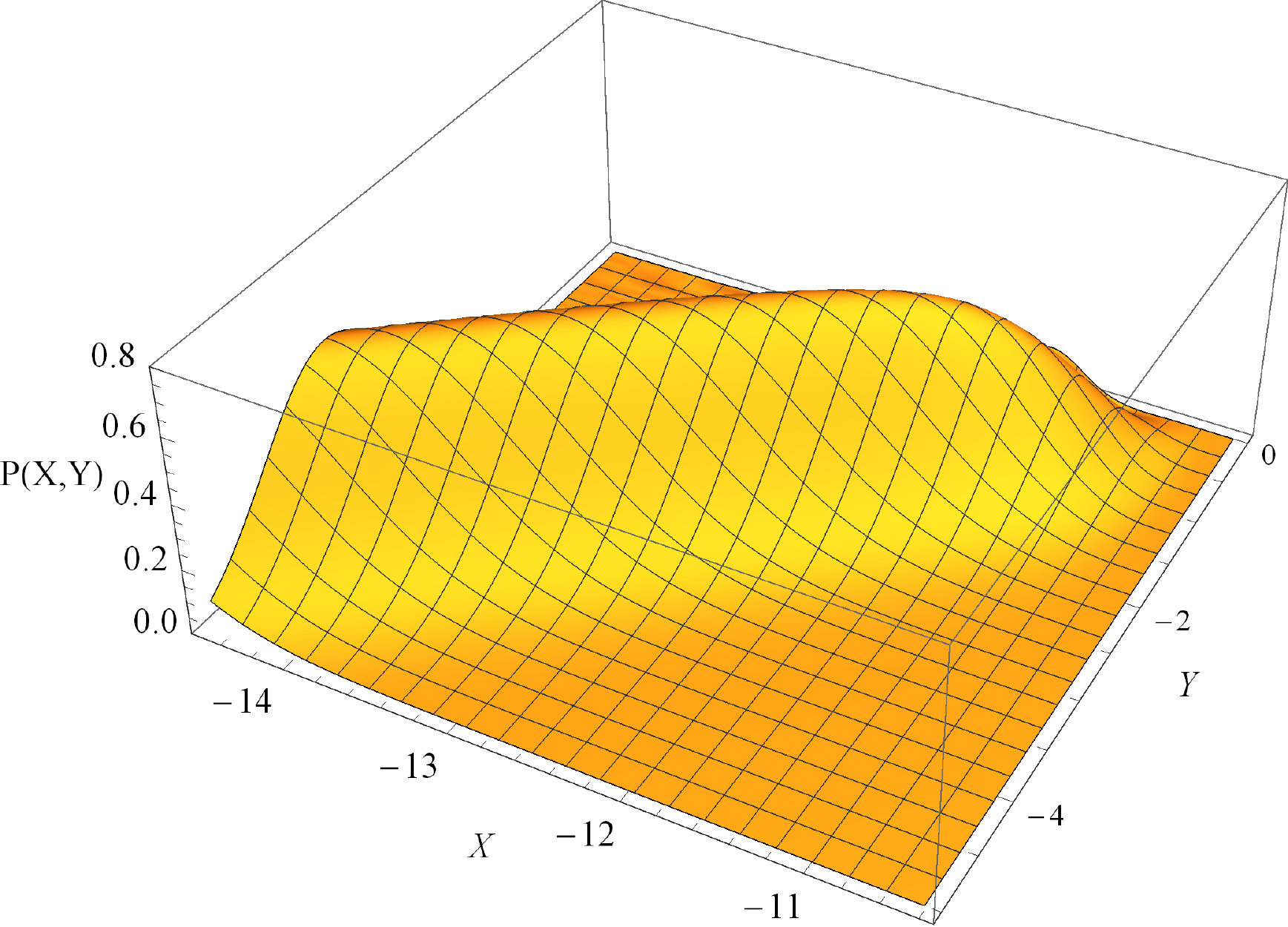}
    \caption{Left: The black line is the on-shell trajectory Eq.~(\ref{eq:XY}) and curve (D) in Fig.~\ref{fig:00}. Right: the modulus squared of the wave function correspondingly is shown numerically with $r_s=A=1$ and $\sigma=0.75$.
    Here, the domain is set $\left(\widetilde{X}_{L},\widetilde{X}_{R}\right)$ × $\left(-\widetilde{Y}_{m},\widetilde{Y}_{M}\right)=\left(-16, -10\right)$ × $\left(-5, 0.5\right)$. The incoming pulse from $X,$ $Y\rightarrow -\infty$ (corresponding to the cosmological $r_c$) decays at the potential barrier boundary. In Appendix~\ref{appendix:out}, we show that the wave is indeed bounded due to the potential barrier.} 
    \label{fig:02}
\end{figure}

{\color{black} Before analyzing the interpretation of the wave function, we must comment on the boundary conditions. The choice of the boundary condition $\Psi(\widetilde{X}_{R}, Y)=0$ is sensitive to the $\sigma$. Since the classical trajectory (D) crosses the potential barrier and is near the potential hollow, the broad Gaussian wave packet, a large value of the $\sigma$, may cross the potential hollow and become unbounded. In Fig.~\ref{fig:02}, the wave function would be unbounded once we set the boundary at $\widetilde{X}_{R}=-9$. However, it is just an ill-defined $\sigma$. If we consider a Gaussian wave with $\sigma=0.1$, the wave function is no longer unbounded even if we choose a larger boundary of $\widetilde{X}_{R}$.}

{\color{black}In Fig.~\ref{fig:02}, the steepest-descent still locates on the classical on-shell trajectory Eq.~(\ref{eq:XY}) and the trajectory crosses the effective potential barrier boundary. In other words, if an incoming pulse is sent at the cosmological horizon $r_c$, it vanishes as soon as it crosses the effective potential barrier boundary. Since the cosmological constant has the Nariai limit $\Lambda\leq4/9$ (to preserve horizons), the nonvanishing wave packet is in the interval,
\begin{eqnarray}\label{eq:ax}
r_c\left(=0\right)\le{a^2}(t)\leq e^{2X_{b}}\lesssim r_c+M\left(=1\right)\vert_{\Lambda=4/9},
\end{eqnarray}
where $X_b$ is the coordinate that the wave crosses the effective potential barrier boundary. The boundaries of the nonvanishing wave packet in the Penrose diagram are shown in Fig.~\ref{fig:00PD}. For $\Lambda=10^{-9}$, $X_b\sim -11$ as shown in the left Fig.~\ref{fig:02}.

Since the spacetime outside the cosmological horizon is purely classical, the steepest descent, surprisingly, is not clear beyond the potential barrier boundary. This may imply that \textit{the Schwarzschild-de Sitter spacetime loses the classical meaning beyond the $r \sim r_c+M$ hypersurface.} Besides, this also implies a compelling interpretation that the DeWitt boundary condition avoids \textit{infinity} in the $r\rightarrow\infty$ limit in metric Eq.~(\ref{eq:sbh}). Interestingly, the $X\sim 0$ hypersurfaces, $r\sim M$ and $r\sim r_c+M$, not only share the same metric form in Eq.~(\ref{eq:sbh}) but also satisfy the DeWitt boundary condition.}

\section{Conclusion}\label{sec:con}
In this work, we solve the Wheeler-DeWitt equation for the Schwarzschild-(anti) de Sitter metric. Since the equation is impossible to solve simply by separating variables, we use a numerical approach. We analyze the wave function inside the black hole horizon (for $\Lambda<0$, $\Lambda=0$, $\Lambda>0$) and outside the cosmological horizon (for $\Lambda>0$).

The wave function solution of the Wheeler-DeWitt equation depends on the potential with an additional $\Lambda$ term. The sign of $\Lambda$ determines the potential behavior. When $\Lambda\le0$, the potential behaves like a potential hollow. When $\Lambda>0$, the potential hollow dwindles and a potential barrier appears. In Fig.~\ref{fig:01}, we show that the  \textit{annihilation-to-nothing} is a generic scenario even with the existence of $\Lambda$, and the DeWitt boundary condition, vanishing boundary condition, yields at $X\sim0$ ($r \sim M$)~\cite{Bouhmadi-Lopez:2019kkt,Brahma:2021xjy}. The steepest descent locates on the classical on-shell trajectory well as expected. 

In de Sitter spacetime, the Schwarzchild-de Sitter metric is valid both inside the black hole horizon and outside the cosmological horizon. Therefore, the Wheeler-DeWitt equation for this metric can be extended to the region outside the cosmological horizon. The wave function on the classical on-shell trajectory vanishes at the effective potential barrier boundary. It means that there is no clear classical interpretation beyond this boundary. The Schwarzchild-de Sitter metric may not be able to describe regions far away from the cosmological horizon. The conservative interpretation is that the DeWitt boundary condition avoids infinity in the $r\rightarrow\infty$ limit in metric Eq.~(\ref{eq:sbh}). The radical interpretation is that the $r\sim r_c+M$ hypersurface is the furthest spacetime that the Schwarzchild-de Sitter metric can reach.

It is interesting to mention that we can provide the vanishing boundary condition beyond the cosmological horizon. If this is the case, the natural consequence is that we will lose the quantum coherence as the spacelike hypersurface curves beyond the cosmological horizon. What does this mean? This goes beyond the scope of this paper. However, one natural consequence is this: it is not surprising to see the decoherence of quantum modes that go beyond the cosmological horizon. This is deeply related to the origin of structures of our universe in the inflationary universe. Our approach might be an alternative explanation to the question of why quantum fluctuations are frozen to classical perturbations, while this issue is deeply related to the decoherence \cite{Kiefer:1998qe}. We leave this interesting research topic for future investigations.

\appendix
\section{The Boundedness of the Wave Function Inside the Black Hole Horizon}\label{appendix:unb}
\begin{figure}[H]
    \centering
    \includegraphics[width=0.45\textwidth]{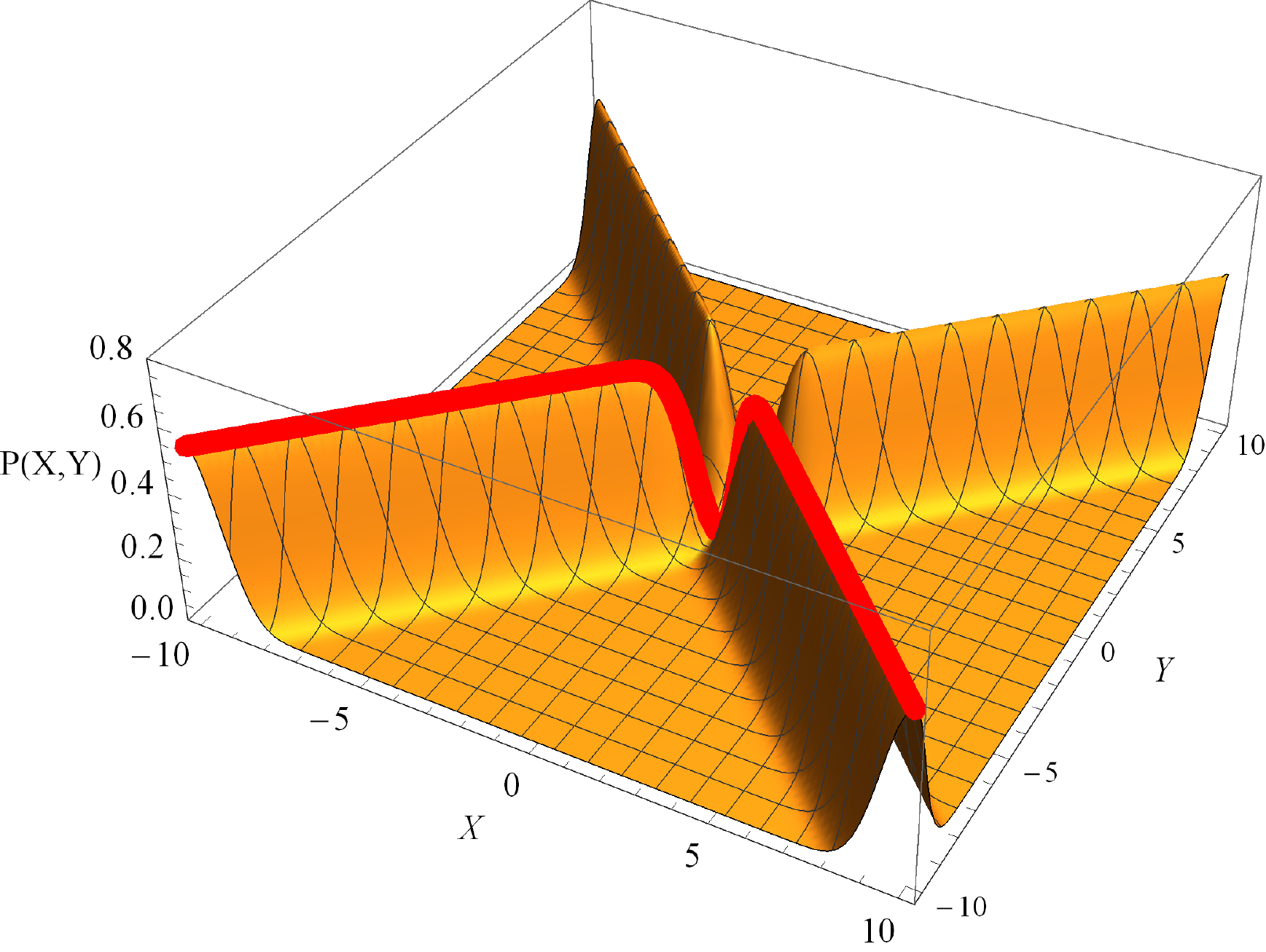}
    \caption{The modulus squared of the wave function is shown numerically with $r_s=0$, $\Lambda=A=1$, and $\sigma=0.75$.}
    \label{fig:a01}
\end{figure}

To understand the origin of the unboundedness, we first turn off the potential in Eq.~(\ref{eq:V}) by setting $r_s=0$ as shown in Fig.~\ref{fig:a01}. With the boundary condition set in Sec.~\ref{sec:IN} and the domain $\left(X_L, X_R\right)$ × $\left(-Y_m, Y_M\right)=\left(-10, 10\right)$ × $\left(-10, 10\right)$, we find two extra pulses in $+Y$ region.

When we turn on the potential, with $r_s=1$ and $\Lambda=0$, the amplitude of these extra pulses grows exponentially in the $+Y$ region, and it becomes unbounded because the potential behaves like a potential hollow as explained at the end of Sec.~\ref{sec:WF}. 

\begin{figure}[!ht]
    \centering
    \includegraphics[width=0.3\textwidth]{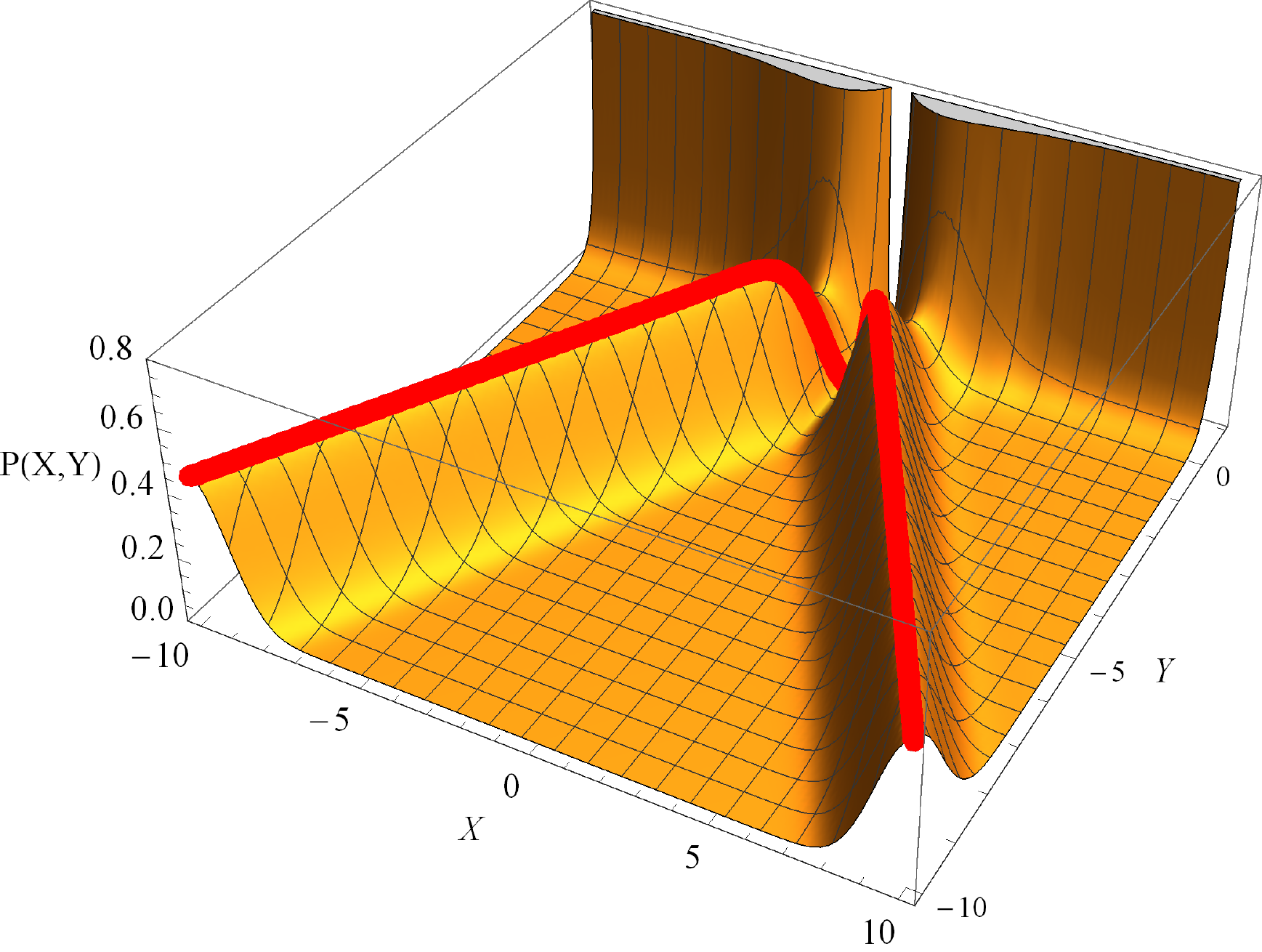}\hspace{0.05 cm}
    \includegraphics[width=0.3\textwidth]{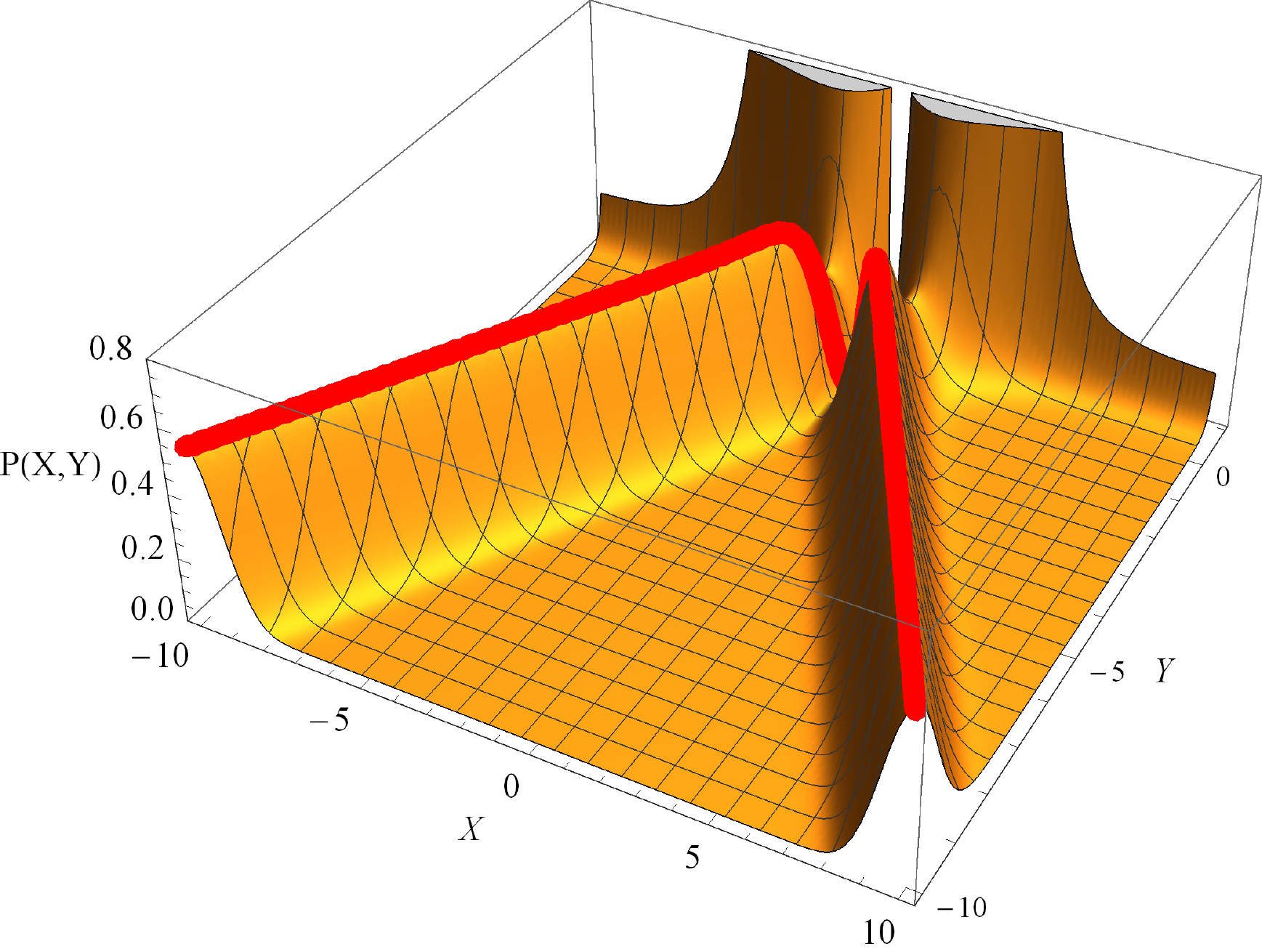}\hspace{0.05 cm}
    \includegraphics[width=0.3\textwidth]{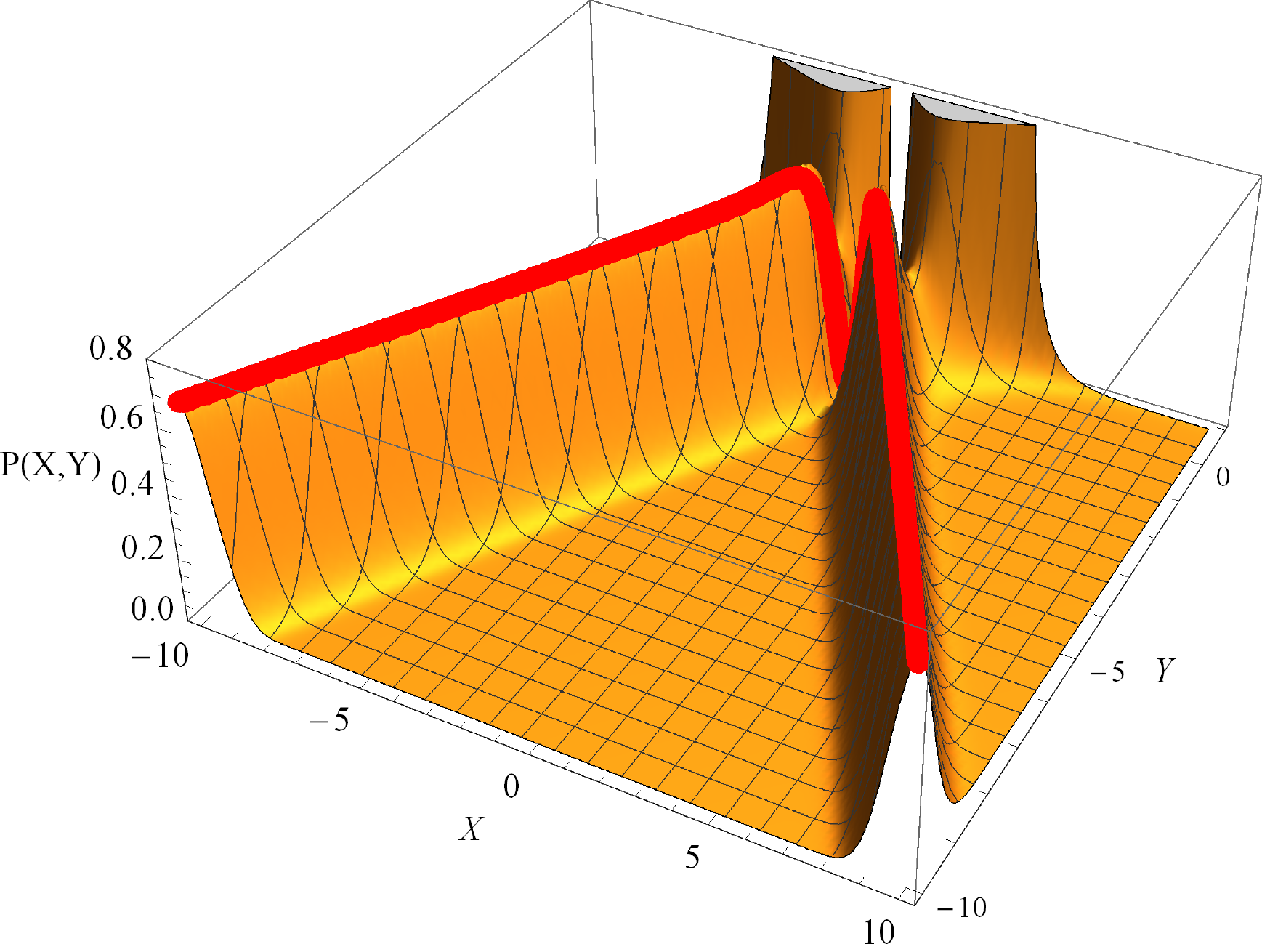}
    \caption{The modulus squared of the wave function is shown numerically for $\sigma=0.9$, $\sigma=0.75$, and $\sigma=0.6$, from left to right, respectively, with $r_s=A=1$ and $\Lambda=0$.} 
    \label{fig:as}
\end{figure}

We have shown that two wave packets in the $-Y$ region are annihilated; the unbounded behavior occurs because the pulses reach the region where the potential is significant. As we can see in Fig.~\ref{fig:as}, as the pulses are localized (equivalently, as $\sigma$ decreases), the unbounded part disappears more and more. In the extreme limit, it will be consistent with the analytic solutions, where the wave function is bounded, as shown in Fig.~(3) in Ref.~\cite{Bouhmadi-Lopez:2019kkt}. Therefore, we can conclude that our results are sufficiently bounded as we carefully choose the boundary conditions.

\section{The Boundedness of the Wave Function Outside the Cosmological Horizon}\label{appendix:out}
\begin{figure}[!ht]
    \centering
    \includegraphics[width=0.45\textwidth]{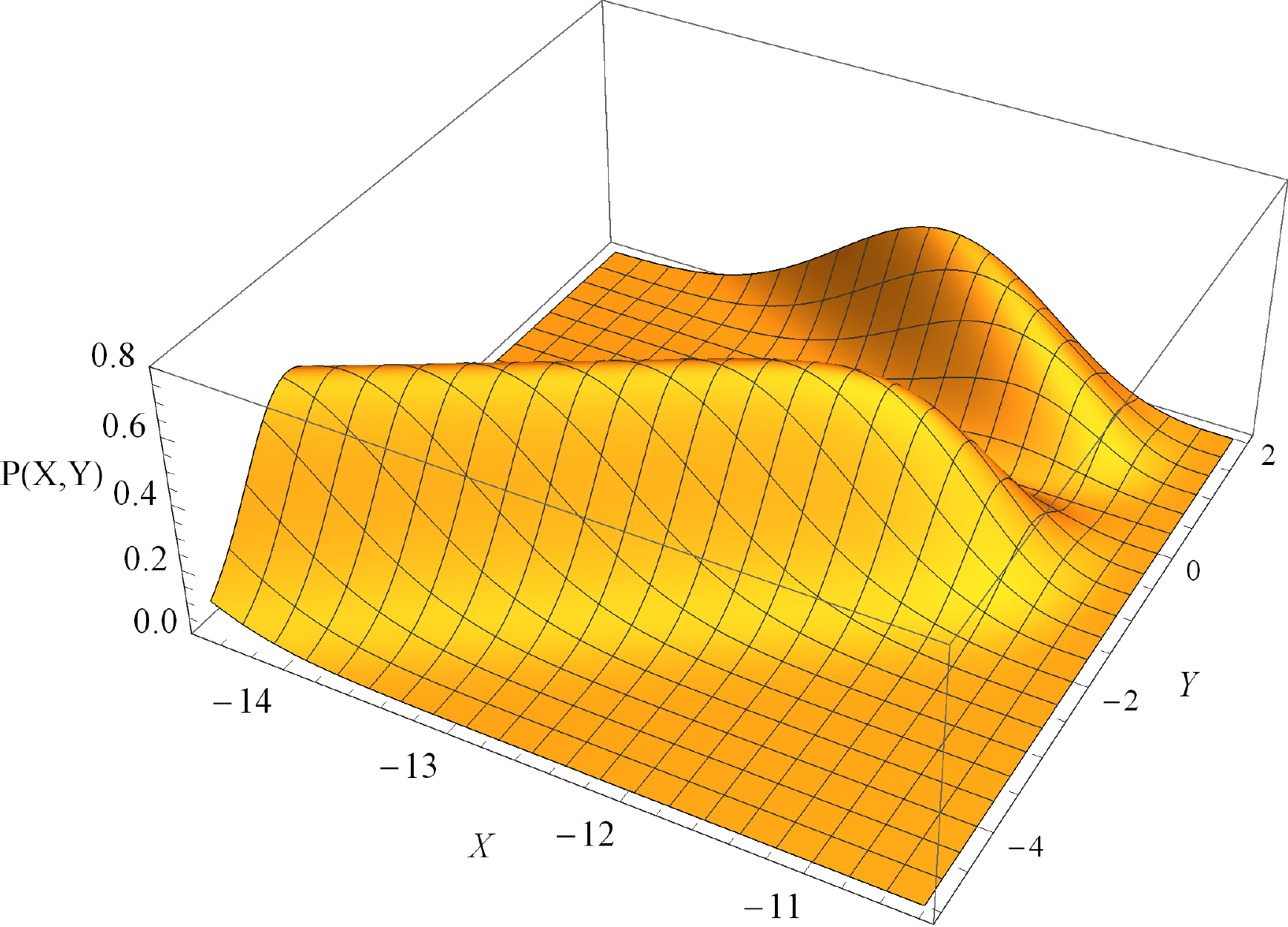}\hspace{0.05 cm}
    \includegraphics[width=0.45\textwidth]{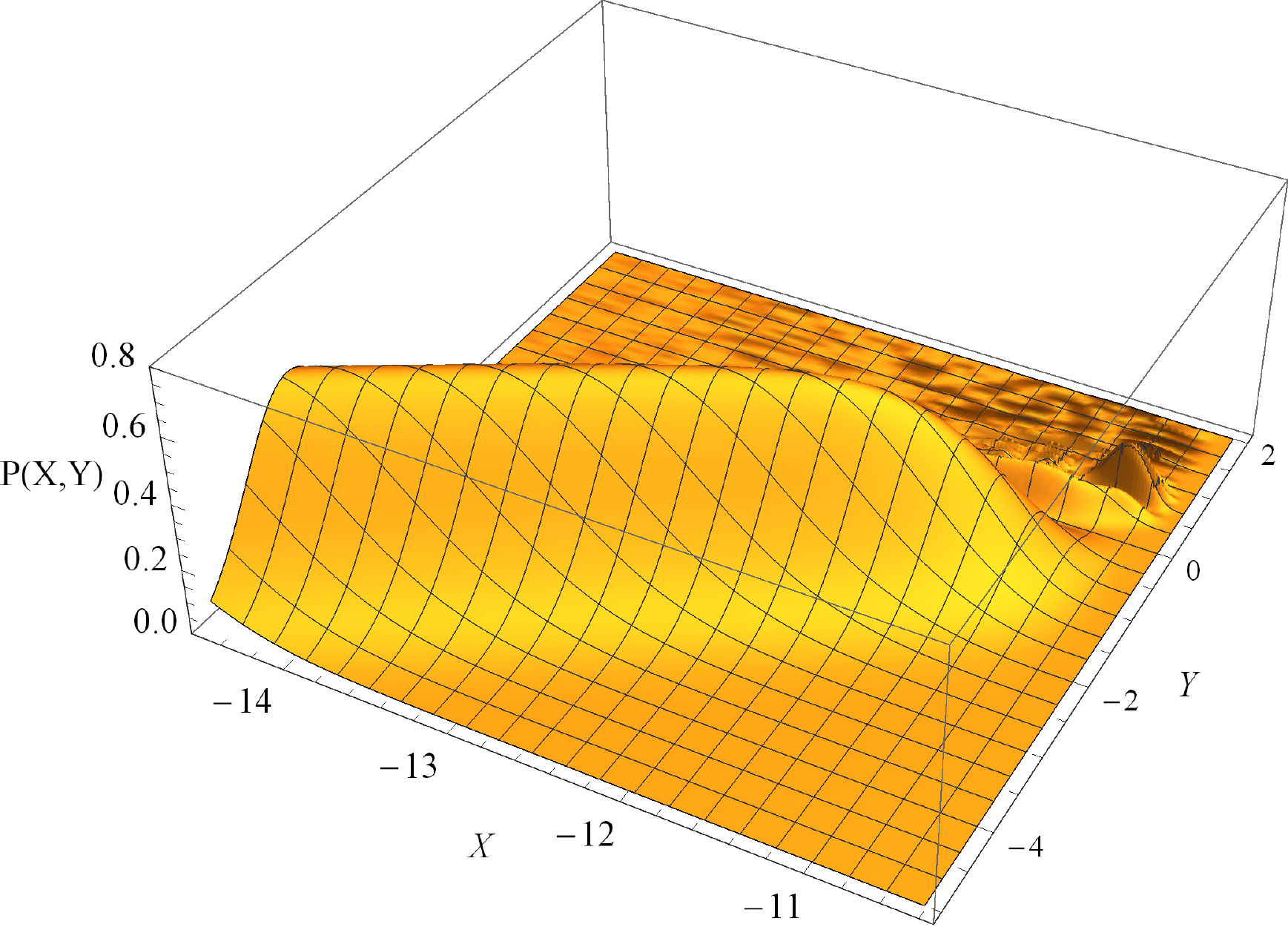}
    \caption{The modulus squared of the wave function without (left) and with potential (right), $r_s=0$ and $r_s=1$, respectively. Here, we set $\sigma=A=1$ and $\Lambda=10^{-9}$.} 
    \label{fig:b1}
\end{figure}

We first compare results with/without the potential Eq.~(\ref{eq:V}) by setting $r_s=0$ / $r_s=1$. With the boundary condition set in Sec.~\ref{sec:OT} and the domain $\left(\widetilde{X}_L, \widetilde{X}_R\right)\times\left(-\widetilde{Y}_m, \widetilde{Y}_M\right)=\left(-16, -10\right)$ × $\left(-5, 2\right)$, we find a pulse in $+Y$ region. Such the pulse is eliminated by the potential barrier, however, there remain some fluctuations that can be interpreted as numerical errors, see Fig.~(\ref{fig:b1}).

\begin{acknowledgements}
We thank Che-Yu Chen for his constructive comments which helped to improve the quality of the manuscript. CHC is supported by the Institute of Physics in Academia Sinica, Taiwan. GT is supported by Basic Science Research Program through the National Research Foundation of Korea, funded by the Ministry of Education (Grant No. 2022R1I1A1A01053784, No. 2021R1A2C1005748). DY is supported by the National Research Foundation of Korea (Grant No. 2021R1C1C1008622, No. 2021R1A4A5031460).
\end{acknowledgements}

\end{document}